\documentclass[twocolumn]{aastex631}
\usepackage{CJK}

\newcommand{\Hi}{H~{\sc i}}

\newcommand{\Cii}{C~{\sc ii}}
\newcommand{\Civ}{C~{\sc iv}}
\newcommand{\Oi}{O~{\sc i}}
\newcommand{\Siii}{Si~{\sc ii}}

\newcommand{\Siiv}{Si~{\sc iv}}

\newcommand{\zsys}{$z_{\text{sys}}$}
\newcommand{\Lya}{Ly$\alpha$}

\newcommand{\kms}{\ifmmode\,{\rm km}\,{\rm s}^{-1}\else km$\,$s$^{-1}$\fi}
\newcommand{\Msun}{\ifmmode\,{\rm M}_{\odot}\else M$_{\odot}$\fi}
\newcommand{\Msunyr}{\ifmmode\,{\rm M}_{\odot}\,{\rm yr}^{-1}\else M$_{\odot}$\,yr$^{-1}$\fi}

\begin{document}
\begin{CJK*}{UTF8}{gbsn}

\title{Detection of gas inflow during the onset of a starburst in a low-mass galaxy at z=2.45}

\shorttitle{Gas inflow in a low-mass $z\sim2.5$ galaxy}

\shortauthors{Coleman et al.}

\correspondingauthor{Erin Coleman}
\email{coleman@gustavus.edu}

\author[0009-0008-3927-4638]{Erin Coleman}
\affiliation{Department of Physics, Gustavus Adolphus College, 
800 W College Avenue, Saint Peter, MN 56082, USA}

\author[0000-0002-2645-679X]{Keerthi Vasan G.C.}
\affiliation{Department of Physics and Astronomy, University of California Davis, 1 Shields Avenue, Davis, CA 95616, USA} 

\author[0000-0003-4520-5395]{Yuguang Chen (陈昱光)}
\affiliation{Department of Physics and Astronomy, University of California Davis, 1 Shields Avenue, Davis, CA 95616, USA} 

\author[0000-0001-5860-3419]{Tucker Jones}
\affiliation{Department of Physics and Astronomy, University of California Davis, 1 Shields Avenue, Davis, CA 95616, USA}

\author[0009-0007-0184-8176]{Sunny Rhoades}
\affiliation{Department of Physics and Astronomy, University of California Davis, 1 Shields Avenue, Davis, CA 95616, USA}

\author[0000-0001-7782-7071]{Richard Ellis}
\affiliation{Department of Physics \& Astronomy, University College London. Gower St., London WC1E 6BT, UK}

\author{Dan Stark}
\affiliation{Steward Observatory, University of Arizona, 933 N Cherry Avenue, Tucson, AZ 85721, USA}

\author[0000-0003-4570-3159]{Nicha Leethochawalit}
\affiliation{National Astronomical Research Institute of Thailand (NARIT), Mae Rim, Chiang Mai, 50180, Thailand}

\author[0000-0003-4792-9119]{Ryan Sanders}
\affiliation{Department of Physics and Astronomy, University of Kentucky, 505 Rose Street, Lexington, KY 40506, USA}

\author{Kris Mortensen}
\affiliation{Department of Physics and Astronomy, University of California Davis, 1 Shields Avenue, Davis, CA 95616, USA} 

\author[0000-0002-3254-9044]{Karl Glazebrook}
\affiliation{Centre for Astrophysics and Supercomputing, Swinburne University of Technology, Hawthorn, Victoria 3122, Australia}
\affiliation{ARC Centre of Excellence for All Sky Astrophysics in 3 Dimensions (ASTRO 3D), Australia}

\author[0000-0003-1362-9302]{Glenn G. Kacprzak}
\affiliation{Centre for Astrophysics and Supercomputing, Swinburne University of Technology, Hawthorn, Victoria 3122, Australia}
\affiliation{ARC Centre of Excellence for All Sky Astrophysics in 3 Dimensions (ASTRO 3D), Australia}



\begin{abstract}


The baryon cycle is crucial for understanding galaxy formation, as gas inflows and outflows vary throughout a galaxy's lifetime and affect its star formation rate.
Despite the necessity of accretion for galaxy growth at high redshifts, direct observations of inflowing gas have proven elusive especially at $z\gtrsim2$.
We present spectroscopic analysis of a galaxy at redshift $z=2.45$ which exhibits signs of inflow in several ultraviolet interstellar absorption lines, with no clear outflow signatures. The absorption lines are redshifted by $\sim$250 \kms\ with respect to the systemic redshift, and \Civ\ shows a prominent inverse P-Cygni profile. Simple stellar population models suggest that this galaxy has a low metallicity ($\sim$5\% solar), with a very young starburst of age $\sim$4 Myr dominating the ultraviolet luminosity. 
The gas inflow velocity and nebular velocity dispersion suggest an approximate halo mass of order $\sim 10^{11}M_{\odot}$, a regime in which simulations predict that bursty star formation is common at this redshift.
We conclude that this system is likely in the beginning of a cycle of bursty star formation, where inflow and star formation rates are high, but where supernovae and other feedback processes have not yet launched strong outflows. 
In this scenario, we expect the inflow-dominated phase to be observable (e.g., with net redshifted ISM absorption) for only a short timescale after a starburst onset. 
This result represents a promising avenue for probing the full baryon cycle, including inflows, during the formative phases of low-mass galaxies at high redshifts.

\end{abstract}



\section{Introduction} \label{sec:intro}

The growth of galaxies is driven by gravitational accretion, a process that channels intergalactic and circumgalactic gas into galaxies, providing the essential material for star formation \citep[e.g.,][]{tumlinson_2017}. Gas inflow is a critical component of the baryon cycle, which underscores the continuous import of gas as a fundamental driver of galactic evolution and star formation rates. The gas accretion is, however, modulated by significant outflows of gas driven by star formation and supermassive black hole feedback \citep[e.g.,][]{murray_2005,hopkins_2012}. 
The efficiency and scale of gas inflows and outflows are closely coupled to the gravitational potential, and they can vary significantly among different galaxies and over their lifespans \citep[e.g.,][]{somerville_dave_2015}. 

Despite the critical role of gas accretion in galactic evolution, observational signatures of accretion have proven elusive, especially at high redshifts. 
Star-forming galaxies at $z \gtrsim 2$ instead almost ubiquitously exhibit strong \textit{outflows} \citep[e.g.,][]{shapley_2003, steidel_structure_2010,pahl_2020,weldon_mosdef-lris_2023}. The presence of outflowing gas is most commonly detected via blueshifted interstellar (IS) absorption lines and/or redshifted Lyman-alpha (\Lya) emission in ``down-the-barrel'' (DTB) spectra of galaxies. These spectral features are widely interpreted as indicators of gas being transported by large-scale, volume-filling galactic outflows driven by star formation and supermassive black hole feedback \citep[e.g.,][]{fg23}. 
Gas inflows are expected to show redshifted absorption and blueshifted resonant emission in DTB spectra, opposite to the P-Cygni outflow profile, yet it is clear that outflowing gas dominates this signature in the vast majority of star-forming galaxies studied to date (with only a few percent being dominated by inflows at $z\sim2$; \citealt{weldon_mosdef-lris_2023}). 
These outflows are believed to contribute significantly to the circumgalactic medium (CGM), with observations showing that the majority of heavy elements produced by star formation now reside outside of galaxies (in their CGM or beyond; e.g., \citealt{peeples_2014,jones_2018,sanders_2021}). 

The lack of observed inflows toward galaxies may partly arise from a low covering fraction of the accreting gas. Theoretical work suggests that cold accretion via filamentary inflows can channel gas through narrow pathways into galaxies, resulting in limited coverage in DTB spectra \citep[e.g.,][]{dekel_cold_2009, stewart11}. Statistical analyses of background galaxy sightlines additionally reveal that cool gas ($T \lesssim 10^4$~K) within the virial radius ($r_\mathrm{vir}$) predominantly comprises kinematically outflowing material \citep{chen20}, highlighting the observational bias toward outflows viewed down-the-barrel. 
Observational evidence of inflows at high redshifts has instead come from spatially extended \Lya\ emission in the CGM, tracing cool gas which may indeed form narrow streams \citep[e.g., ][]{vanzella17, martin19}. 
Yet despite the paucity of observational signatures, inflows must nonetheless dominate over outflows in terms of total mass flux in order for galaxies to grow more massive over time. 

While outflows statistically dominate observed CGM kinematics, the outflow, inflow, and star-formation rates of galaxies vary substantially across cosmic time. The peak in cosmic star formation density occurs at $z \sim 2$--3 \citep{madau2014}, coinciding with galaxies having large gas fractions and high accretion rates \citep[e.g.,][]{muratov_gusty_2015,tacconi_2020}. Consequently, galaxies at this epoch can undergo intense and highly time-variable ``bursty'' star formation \citep{atek2014,sparre_starbursts_2017}. The phenomenon of bursty star formation is especially prominent in dwarf galaxies (stellar mass $M_* \lesssim 10^9 M_\odot$), where shallow gravitational potentials are more affected by stellar feedback, leading to cyclical star formation patterns \citep[e.g.,][]{guo16}. 
Recent observations at $z\gtrsim6$ have further highlighted the importance of bursty star formation, which may partly explain the number density of galaxies with high UV luminosity \citep{finkelstein22, harikane23} as well as galaxies exhibiting post-starburst stellar populations \citep{carnall_2023,looser_2024,langeroodi_2024}. This underscores the importance of understanding galaxy-gas interactions in the bursty star formation regime. 
Notably for this work, bursty star formation patterns may potentially allow observations of inflow-dominated CGM kinematics during periods between starbursts, when galactic-scale outflows are absent.

While bursty star formation is expected to be most prevalent in low-mass and high-redshift galaxies, such sources are challenging to observe spectroscopically due to their faintness (typically observed optical magnitudes $\gtrsim 25$ for galaxies with stellar mass $M_* \lesssim 10^9 M_\odot$ at $z \simeq 2.5$). However, magnification from strong gravitational lensing can provide a valuable means to probe their physical properties \citep[e.g.,][]{richard_2011,stark_2014,snapp-kolas_2023,vasan_2023,vasan_2024}.

In this paper we study a low-mass galaxy at $z=2.45$ which is strongly lensed by a foreground galaxy group at $z = 0.214$. The lens system, identified in the Cambridge Sloan Survey of Wide Arcs in the Sky \citep[CSWA;][]{belokurov_2009} with a designated ID of CSWA 128, was initially recognized due to a bright strongly lensed source at $z = 2.225$ \citep{stark_cassowary_2013}. The $z = 2.45$ source was first identified from adaptive optics imaging by \citet{sharma_high_2018}, who estimated its redshift as $z \simeq 2.9$, although our spectroscopic analysis reveals a redshift of $z=2.45$.
In their work, the source was labeled System 10. For consistency, we will also refer to this $z = 2.45$ galaxy as System 10 throughout this study. Figure \ref{f:galaxyimage} shows the foreground $z=2.225$ source and System 10.
The galaxy has a faint intrinsic magnitude $g = 27.7\pm0.2$ but has three highly magnified images with observed $g \lesssim 25.5$ \citep[see Section~\ref{sec:luminosity} and][for discussion of photometry and lensing magnification]{sharma_high_2018}, making spectroscopic observations feasible. 
As we describe herein, this galaxy shows prominent spectroscopic evidence of inflowing gas from an inverse P-Cygni interstellar \Civ\ profile and other features, with no clear outflow signatures.

The structure of this paper is as follows. In Section \ref{l:data} we discuss the spectroscopic data collection and processing. 
Section \ref{l:analysis} describes our analysis, which includes establishing the systemic redshift, measuring the velocity profile of several interstellar absorption lines, and modeling the stellar continuum to determine age and metallicity. 
In Section \ref{l:discussion} we discuss our findings in the context of previous observational work and theoretical predictions. Finally, we discuss our conclusions and prospects for future study in Section \ref{l:conclusions}. We adopt a $\Lambda$CDM cosmology with $H_0 = 70~\mathrm{km~s}^{-1}~\mathrm{Mpc}^{-1}$, $\Omega_\Lambda=0.7$, and $\Omega_m = 0.3$. Magnitudes are reported in the AB system \citep{oke_1983}.

\section{Data} \label{l:data}

\begin{figure*}[htb]

\includegraphics[width=0.47\linewidth]{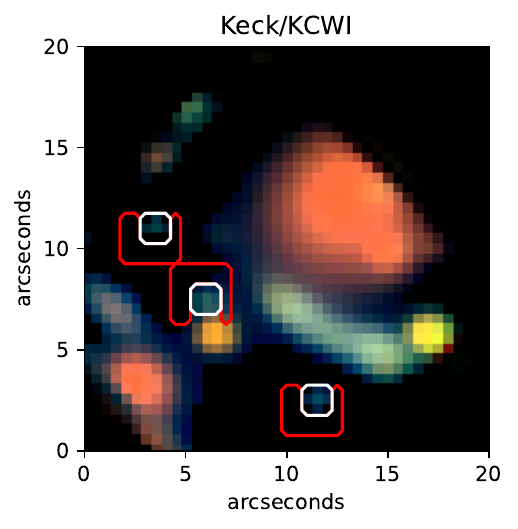}
\includegraphics[width=0.49\linewidth]{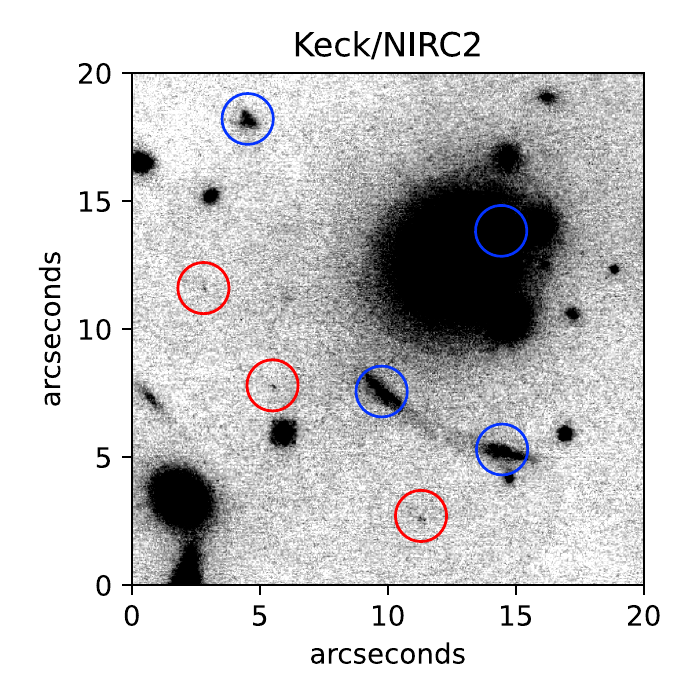}

\caption{Images of the CSWA 128 lens system. Both panels are oriented with north up and east to the left. 
Three magnified images of System 10 are marked in each panel. \textit{Left panel:} Optical false-color image obtained by averaging the KCWI datacube used in this work over three wavelength ranges (blue: $\lambda = 3500$--4100~\AA, green: $\lambda = 4200$--4800~\AA, red: $\lambda = 4900$--5500~\AA). White outlines show the regions used to extract spectra for each image of System 10, and red outlines indicate the regions used to correct sky subtraction residuals around each image. 
\textit{Right panel:} Keck NIRC2 near-IR adaptive optics image (see \citealt{sharma_high_2018} for details). Images of system 10 ($z=2.45$) are circled in red and images of the $z=2.225$ foreground galaxy are circled in blue. The target galaxy images are spatially compact despite the lensing magnification. 
}
\label{f:galaxyimage}
\end{figure*}

CSWA 128 was observed with the Keck Cosmic Web Imager (KCWI; \citealt{morrissey_keck_2018}) on 2 June 2019 and 19-20 June 2020 for a total integration time of 8.2 hours. We used the Medium slicer and blue low-resolution (BL) grating with a central wavelength of 4500~\AA, providing a 16$\times$20 arcsec field of view with wavelength coverage 3500-5500~\AA\ and spectral resolution 2.4~\AA\ FWHM (R~$\simeq 1800$). Data were taken in orthogonal position angles of 45 and 135 degrees in order to achieve an approximately circular point spread function in the combined data cube. 
Sky conditions were clear to partly cloudy (up to $\sim$0.5 mag extinction), with seeing ranging from 0\farcs7--1\farcs25 FWHM. The data were reduced following the procedure described in \cite{mortensen_kinematics_2021}. In short, we used the KDERP-v1.0.2 pipeline which removes instrument signatures and performs sky subtraction, wavelength calibration, and spatial rectification, including a correction for differential atmospheric refraction.  The reduced datacubes were fit with a 2D first order polynomial to correct for residual structure. Individual exposures were then aligned and median combined to produce the final datacube used in this work.
Figure~\ref{f:galaxyimage} shows a false-color image obtained by summing the reduced datacube in three broad wavelength ranges spanning 3500 -- 5500~\AA.

\begin{figure*}[htb]
\begin{center}
\scalebox{.48}{\includegraphics{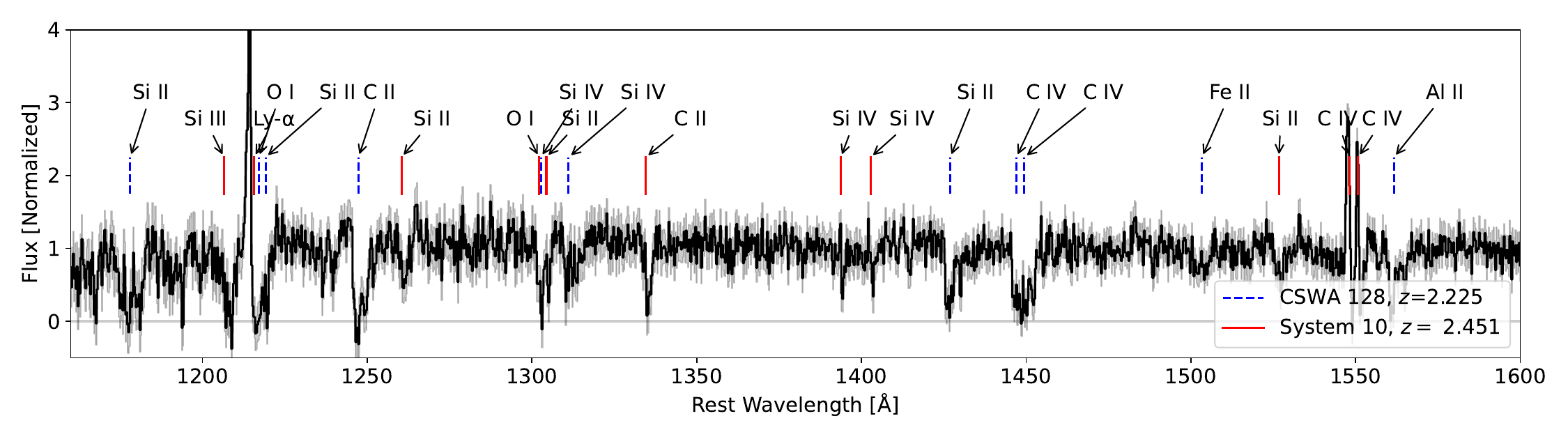}}
\scalebox{.6}{\includegraphics[]{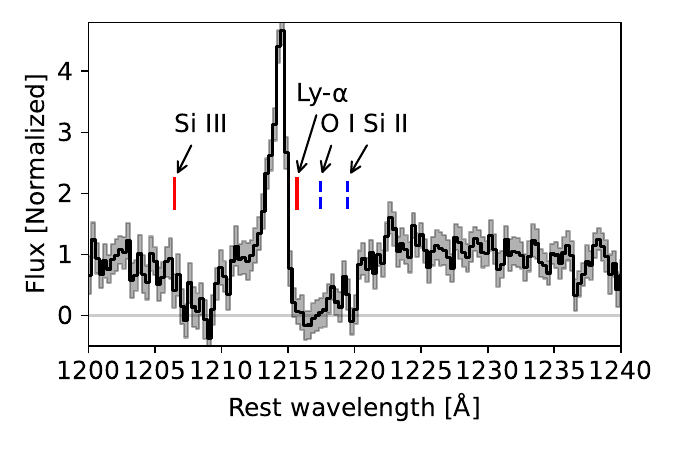}}
\scalebox{.55}{\includegraphics{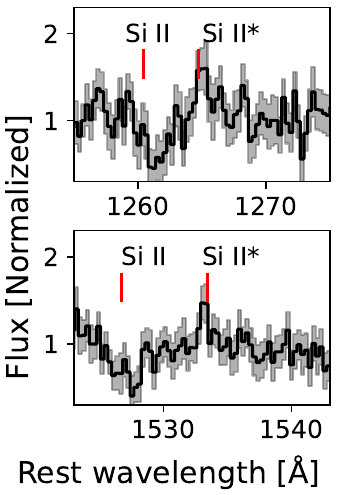}}
\scalebox{.6}{\includegraphics{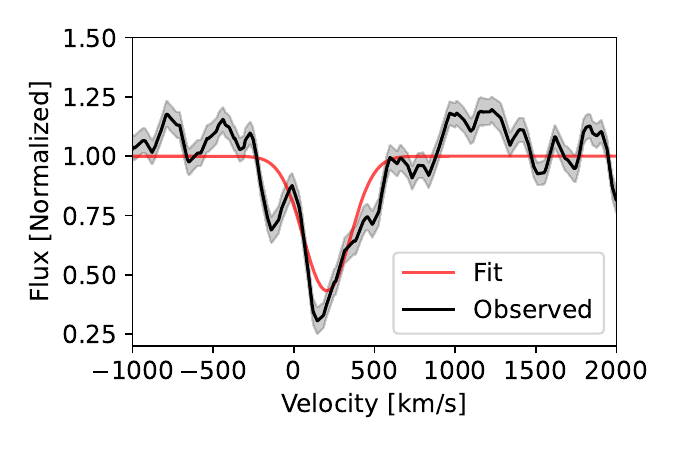}}

\caption{\textit{Top:} Rest-frame spectrum of System 10, obtained by summing the three multiple images. Solid red lines indicate strong features at the redshift of System 10, including ISM absorption which shows signatures of gas inflow. Dashed blue lines indicate intervening absorption features at the redshift $z=2.225$ of the brighter arc, arising from its circumgalactic medium. Spectral regions affected by intervening absorption are not used in this work. \textit{Bottom left:} The summed spectrum zoomed in around the \Lya\ feature. The red side of \Lya\ is affected by intervening \Oi~$\lambda$1302 and \Siii~$\lambda$1304 absorption from the foreground arc, such that we cannot reliably determine the \Lya\ emission properties. However there is clearly a component of blueshifted \Lya\ emission, indicative of low \Hi\ column density or inflowing gas. 
\textit{Bottom center:} The summed spectrum zoomed in around the \Siii*~$\lambda \lambda$1264, 1533 fine structure emission lines. The \Siii* $\lambda$1309 line is not used in this work because it is affected by intervening absorption.
\textit{Bottom right:} Average velocity profile of several strong ISM absorption lines with respect to the systemic redshift of the galaxy. The red line shows a best-fit Gaussian profile. The absorption is clearly redshifted, indicating inflowing gas, while the absorption at negative (blueshifted) velocities is nearly zero.
Shaded regions in all panels show the $1\sigma$ uncertainty. 
}

\label{f:fullspectrum}
\end{center}
\end{figure*}

The target of this study is the $z=2.45$ source with three lensed images shown in Figure~\ref{f:galaxyimage}, first identified as a multiply imaged galaxy and denoted System 10 by \cite{sharma_high_2018} based on adaptive optics imaging (right panel of Figure~\ref{f:galaxyimage}). From their lens model, this galaxy was estimated to have a redshift of $z=2.90\pm0.25$ \citep{sharma_high_2018}. Our KCWI data confirm the multiple images with a spectroscopic redshift of $z=2.45$ (Figure~\ref{f:fullspectrum}; Section~\ref{ss:systemicz}). 
The KCWI datacube is affected by variable sky subtraction residuals, resulting from a dearth of blank sky regions in this lensing group field, with amplitude comparable to the flux of our target. We correct for this residual structure using the average of blank sky spaxels near each of the target images. The reference sky regions are shown in Figure \ref{f:galaxyimage}; additional blank sky regions were also used to estimate the uncertainty in our spectra. 
We subtract the adjacent sky residual from the spectrum of each lensed image of System 10, and confirm that the spectra of all three images have consistent shapes. We then sum the spectra of all three images to increase the signal-to-noise ratio. 
The resulting 1-D summed spectrum is shown in Figure~\ref{f:fullspectrum}. 
In the summed spectrum, the continuum is detected with signal-to-noise ratio of 5.7 per spectral pixel at rest-frame 1350~\AA, which is representative of the regions used in this work.
Numerous nebular emission and interstellar absorption lines are detected with high significance at the galaxy's redshift, as well as intervening absorption from the circumgalactic medium of the bright foreground lensed source at $z=2.225$.

\section{Analysis}\label{l:analysis}

System 10 exhibits strong \Civ~$\lambda\lambda$1549, 1551 nebular emission characteristic of young, hot stars \citep[e.g.,][]{senchyna_direct_2022}. The interstellar absorption components of this \Civ\ feature are clearly redshifted compared to the nebular emission components, which is a robust signature that System 10 is experiencing gas inflow \citep[e.g.,][]{carr_2022}. In fact all of the well-detected ISM absorption lines have centroids which are are redshifted with respect to the nebular emission, indicating that the absorbing medium has a positive line-of-sight velocity (i.e., inflowing) relative to the star-forming regions. 
As redshifted absorption is unusual in galaxies at this cosmic epoch (e.g., observed in only 3 out of 134 galaxies by \citealt{weldon_mosdef-lris_2023}; see also \citealt{steidel_structure_2010}), we explore the characteristics of this absorption and offer a theoretically-motivated explanation of this phenomenon. In this section we describe the measurement of systemic redshift, inflowing gas kinematics, and modeling of the young stellar population properties.

\subsection{Systemic redshift} \label{ss:systemicz}

While we detect nebular emission from the \Civ~$\lambda\lambda$1549, 1551 doublet, it does not necessarily trace the systemic redshift due to resonant scattering. Ideally we would use stellar or non-resonant nebular emission lines to determine the redshift, but none are detected in the spectrum of System 10. We instead use fluorescent fine structure emission lines, which have been shown to trace the systemic redshift of $z\sim2$ galaxies within $\lesssim 40$~\kms\ \citep{kornei_2013,jones_2012,prochaska_2011,erb_2012}.
We fit the \Siii*~$\lambda$1264 and $\lambda$1533 emission lines each with a single Gaussian profile plus a linear continuum. 
These lines have signal-to-noise ratios of 5.9 and 4.8, respectively.
We do not attempt to measure \Siii*~$\lambda$1309 due to contamination by intervening absorption at $z=2.225$ (Figure~\ref{f:fullspectrum}). 
Each \Siii* line was fit separately, with free parameters for its redshift, amplitude, and width. The \Siii* emission lines were fit independently from the associated \Siii\ absorption lines, as they are known to have different kinematic profiles.
Table \ref{t:siz} lists the best-fit redshifts for each feature. The two \Siii* lines are in reasonable agreement with a difference of $110\pm70$ \kms, and we adopt their average as our best estimate of the true systemic redshift of System 10: $z_{zys} = 2.4515\pm0.0004$.

We can also estimate the redshift from nebular \Siiv~$\lambda\lambda$1393, 1402 emission. While these are resonant lines, they exhibit less interstellar absorption than \Civ\ and thus are likely to be closer to the true systemic value. We fit the \Siiv\ complex with two Gaussian profiles for the nebular emission, and two Gaussians for the interstellar absorption, all convolved with a Gaussian of 2.4~\AA\ FWHM to account for the instrument line spread function.
The model has four free parameters each for the absorption and emission components: the amplitude of the $\lambda$1393~\AA\ line, the ratio of the 1402~\AA\ amplitude relative to the 1393~\AA\ line, their common line width (i.e., Gaussian $\sigma$), and common redshift. The resulting best-fit emission redshift of $z=2.4514 \pm0.0005$ (Table~\ref{t:siz}) is consistent with $z_{\text{sys}}$ measured from the fine structure lines.

\begin{figure*}[htb]
\begin{center}
 
\scalebox{.75}{\includegraphics{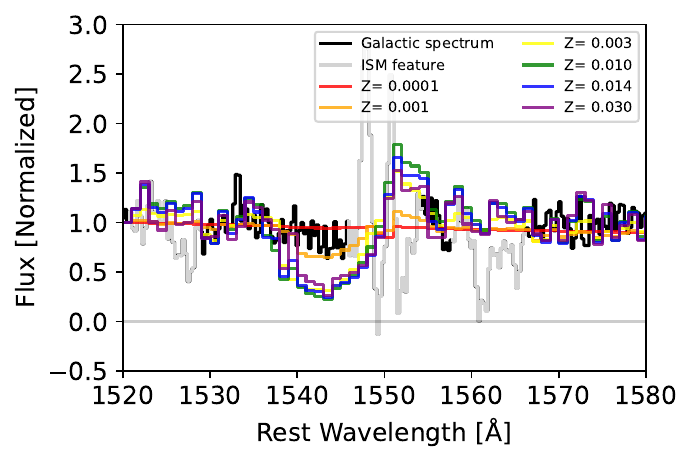}}
\scalebox{.75}{\includegraphics{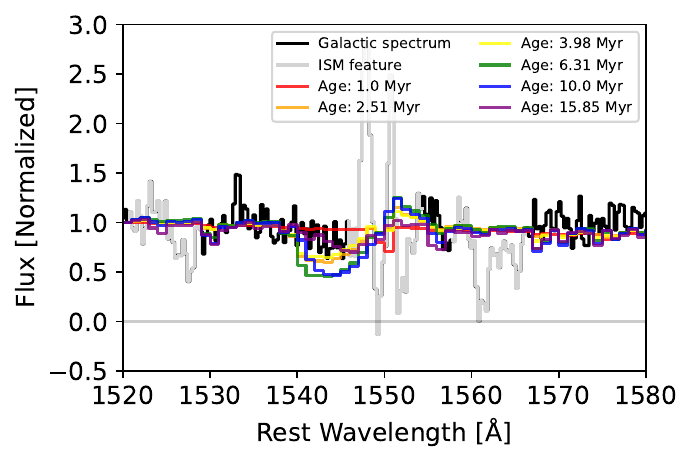}}
\scalebox{.8}{\includegraphics{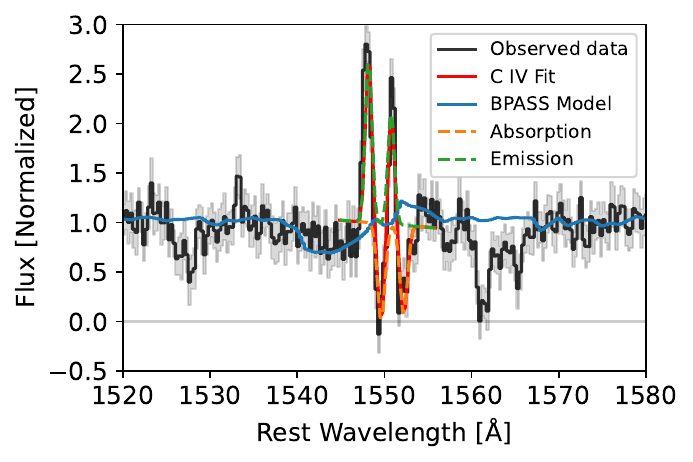}}


\caption{
\textit{Top left:} Stellar continuum models for a simple stellar population showing the effect of varying metallicity (Z) at fixed age. The weak P-Cygni stellar wind absorption component clearly favors a low metallicity ($Z \lesssim 0.001$, or $\lesssim 10$\% solar, for an age of 4 Myr). Regions of the spectrum affected by strong non-stellar features are masked in gray. 
\textit{Top right:} Similar to the center panel, showing the effect of varying the age. For a metallicity $Z=0.001$, the stellar wind absorption indicates ages $\lesssim 5$~Myr. While ages $\gtrsim 15$~Myr can also provide a reasonable fit to the \Civ\ stellar wind, they cannot explain the strong nebular component which requires very massive (thus young) stars. \textit{Bottom:} Spectrum of System 10 in the region of \Civ. Nebular emission is prominent, with interstellar absorption redshifted relative to the emission features, clearly indicative of inflows. The best fit model of the \Civ\ feature is shown in red. This model includes the stellar component, nebular emission, and interstellar absorption. 
}

\label{f:models}
\end{center}
\end{figure*}

We now turn to the \Civ~$\lambda\lambda$1549, 1551 doublet which shows a complex mix of nebular emission, interstellar absorption, and stellar P-Cygni features (Figure~\ref{f:models}). The nebular emission component is a valuable diagnostic of bulk inflow and outflow due to resonant scattering effects, which can broaden the emission lines and cause their peaks to shift. In the case of inflowing gas, the emission component can be scattered to shorter wavelengths. 
We model the total profile as the sum of a stellar continuum template (see Section~\ref{ss:stellarmodels}), and double Gaussian profiles for both the emission and absorption components following the same procedure as for the \Siiv\ doublet. The best-fit total profile is shown in Figure~\ref{f:models}. We find redshifts of the emission component $z_{em} = 2.45062 \pm 0.00008$ and the absorption component $z_{abs} = 2.4547\pm0.0002$.

The \Civ\ emission redshift is indeed lower than $z_{sys}$ corresponding to a blueshift of $-72\pm38$~\kms. We interpret this as the bias from resonant scattering by inflowing gas. Moreover, the blueshifted \Civ\ nebular emission relative to interstellar absorption (Figure~\ref{f:models}) is unambiguous and is clear evidence of inflowing gas. The presence of inflows is therefore a robust conclusion based on the \Civ\ profile, regardless of uncertainty in the precise systemic redshift.

\begin{table}[h!]
\centering
\begin{tabular}{|c| c |} 
 \hline
Line & Redshift \\\hline
\Siii*~$\lambda$1264 Å&
2.4521 $\pm$ 0.0007\\
\Siiv~$\lambda\lambda$1393, 1402 Å&
2.4514 $\pm$ 0.0005\\
\Siii*~$\lambda$1533Å & 2.4508 $\pm$ 0.0005 \\ \hline
\end{tabular}

\caption{The redshifts found by fitting ionized silicon emission lines present in our data. Fluorescent \Siii* lines are expected to trace the systemic redshift, while \Siiv\ may be affected by resonant scattering. In this work we adopt the average of the \Siii* lines as the true systemic redshift, $z_{\text{sys}} = 2.4515\pm0.0004$.}
\label{t:siz}
\end{table}

\subsection{ISM gas kinematics and dynamical mass} \label{ss:ismgas}

We now turn to the gas kinematics, from both nebular emission and ISM absorption, and consider implications for the mass of System 10. In brief, we use nebular emission line widths to estimate dynamical mass within the region of detected emission, while the ISM inflow velocity provides constraints on the total halo mass.
We calculate gas velocities using the systemic redshift found in Section~\ref{ss:systemicz} based on \Siii* emission. 
For ISM absorption, we examined the strongest features covered by the KCWI spectra which include several low- and high-ionization species (labeled in Figure~\ref{f:fullspectrum}). Several strong transitions, most notably \Lya\ (at observed $\lambda \sim 4200$~\AA) and \Oi~$\lambda$1302 + \Siii~$\lambda$1304 (at observed $\lambda \sim 4500$~\AA), are blended with intervening absorption features from the foreground $z=2.225$ arc. These blended features are not used in the analysis of System 10's properties. We thus restrict the ISM absorption analysis to \Siii~$\lambda$1260, \Cii~$\lambda$1334, and \Siii~$\lambda$1526\footnote{We note that higher ionization species \Siiv~$\lambda\lambda$1393, 1402 and \Civ~$\lambda\lambda$1549,1551 show similarly redshifted interstellar absorption but are affected by resonant emission components.}. We fit these unblended lines, again using a Gaussian convolved with the 2.4~\AA\ FHWM instrument line spread function. Table~\ref{t:abslines} lists the velocity dispersions and centroids for each line relative to the estimated systemic $z=2.4515$.

\begin{table}[h!]
\centering
\begin{tabular}{|c| c |c|} 
 \hline
Line & $V_{\text{abs}}$ (\kms)  & $\sigma$ (\kms)\\
     & rel. to $z_{\text{sys}}$ & \\
\hline
\Siii~$\lambda$1260 & $320 \pm 60$ & $340 \pm 120$ \\
\Cii~$\lambda$1334 & $240 \pm 30$ & $170 \pm 40$ \\
\Siii~$\lambda$1526 & $200 \pm 70$ & $220 \pm60$ \\ \hline
\end{tabular}

\caption{The best-fit velocity centroids (relative to systemic redshift \zsys) and velocity dispersions (Gaussian $\sigma$, corrected for KCWI's line spread function) for low ionization absorption lines. 
}
\label{t:abslines}
\end{table}

To examine the ISM velocity structure at higher signal-to-noise, we stacked the low-ionization line profiles (from \Siii~$\lambda$1260, $\lambda$1526, and \Cii~$\lambda$1334). 
We interpolated the spectrum around each of these lines onto a common velocity grid, normalized the continuum around each line, and took the average flux. The resulting average absorption profile is shown in the lower right panel of Figure~\ref{f:fullspectrum}. A best-fit Gaussian profile is also provided, although the dynamics of the absorbing gas may be more complex than can be modeled with a single Gaussian.
From the fit we find a line of sight velocity centroid of $V_{\text{cent}} = 250\pm 30$~\kms\ relative to \zsys\ (excluding $\sim$35~\kms\ uncertainty in \zsys\ itself), and a velocity dispersion $\sigma = 200\pm 40$~\kms. 
As with the \Civ\ profile discussed earlier, the positive velocity of ISM absorption is clearly indicative of inflowing gas. We note that this inflow is traced by, and thus enriched with, heavy elements such as carbon and silicon. These elements must have originated within a galaxy at some previous time. The ISM absorption thus likely arises from a galactic fountain process or possibly accretion from an interacting galaxy.

We did not find significant variation between individual lines or between individual lensed images. The centroid velocities from individual absorption line fits (Table~\ref{t:abslines}) all agree to within two standard deviations. There is no significant variation between lensed images of the system; the absorption velocity centroids of the individual images of System 10 agree within one standard deviation.

A key question is whether the ISM absorption velocity is compatible with gravitationally-driven inflow. We thus consider the terminal accretion velocity (equivalent to the escape velocity) using a NFW \citep[Navarro, Frenk, White:][]{navarro_NFW_1997} halo model. An inflow (or escape) velocity of 250~\kms\ along the line of sight is consistent with a minimum dark matter halo mass of approximately {$M_h \sim 10^{11}$~\Msun.
This is notably lower mass than typical spectroscopic samples of galaxies at cosmic noon \cite[$M_h \sim 10^{12}~\Msun$; e.g.,][]{erb_stellar_2006,kriek_2015, turner_comparison_2017}, commensurate with the small intrinsic size and low luminosity of System 10. We note that this halo mass estimate represents a lower limit for the case of recycling gas, which will not attain the terminal velocity. 

The nebular \Civ\ emission kinematics likewise indicates a relatively low mass for System 10. The best-fit emission line velocity dispersion is $\sigma = 34\pm 6$~\kms, comparable to the lowest dispersions found in the samples of $>$100 cosmic noon galaxies from \cite{erb_stellar_2006} and \cite{price_2016}. Combined with the small size observed in AO imaging, this indicates a substantially lower mass than in typical spectroscopic samples at this epoch. 
The dynamical mass is related to integrated velocity dispersion $\sigma$ by the relation
\begin{equation}
    M_{\mathrm{dyn}} = C\sigma^2r/G,
\end{equation} 
where the virial coefficient $C$ may vary between $\sim$1--5 depending on the mass distribution \citep[e.g.,][]{erb_stellar_2006}. 
For a typical $C\approx3$ \citep{price_2016}, we obtain $M_{dyn} \approx 4\times10^{8}$~\Msun\ within a galactocentric radius $r=500$~pc, which is roughly twice the effective radius (accounting for lensing magnification; Section~\ref{sec:luminosity}).
This is commensurate with the estimate of halo mass based on inflow velocity, as well as abundance matching which suggests that a halo mass of $10^{11} M_{\odot}$ corresponds to stellar mass $\sim 10^8-10^9 M_{\odot}$ \citep[e.g.,][]{behroozi_average_2013}. We note that the stellar mass is likely of order half the dynamical mass within the galaxy, considering the high typical gas fractions at this epoch \citep[e.g.,][]{tacconi_2018,sanders_2023}. 
We do not attempt to measure the stellar mass here given the lack of suitable photometric data.

\subsection{Modelling the stellar spectrum}
\label{ss:stellarmodels}

We fit the observed rest-ultraviolet stellar continuum with Binary Population and Spectral Synthesis (BPASS; \citealt{stanway_re-evaluating_2018}) v2.2 simple stellar population models, in order to estimate the bulk properties of the young stellar population. These models predict the intensity of light emitted at each wavelength by a population of stars with an assumed initial mass function (IMF), age, and metallicity. 
In this case we adopted the \cite{salpeter_1955} IMF with an upper mass cutoff of 300~\Msun, considered to be standard (\texttt{imf135\_300}) by the authors of BPASS. 

We fit the models to the spectrum of System 10 in several wavelength ranges which are sensitive to the stellar age and metallicity. The ranges used correspond to the broad stellar wind features \Civ~$\lambda\lambda$1549, 1551 and \Siiv~$\lambda\lambda$1393, 1402, and between rest-frame 1350--1390~\AA\ which contains photospheric absorption features and anchors the continuum level. These regions were chosen to avoid strong interstellar absorption and emission lines, which are not included in the stellar models. Accordingly the wavelength ranges with narrow nebular emission and ISM absorption in \Civ\ and \Siiv\ were not used in the fit. 
We corrected for the effects of dust reddening by scaling the models with a power-law function in wavelength, $\lambda^{\beta}$, which gives a reasonable approximation and good match to the data over the relatively narrow wavelength range considered here. 
$\beta$ is treated as a free parameter to match the observed continuum slope, with a best-fit value varying with model age and metallicity.
The stellar P-Cygni wind features are highly constraining, whereas we do not resolve individual photospheric lines at the spectral resolution of the KCWI data. 
Our data show weak stellar winds (Figures~\ref{f:fullspectrum}, \ref{f:models}), suggesting a low metallicity \citep[e.g.,][]{senchyna_direct_2022}. 
From a $\chi ^2$ analysis of BPASS models with age ranging between 1 Myr and 40 Myr, and metallicity ranging between 0.00001 and 0.04, we find that the best-fit model has a metallicity $Z=0.001$ ($\sim$5\% of the solar value) and age of 4.0 Myr. This is comparable to results from studies of strong \Civ\ emitters at $z\sim0$. In particular, \cite{senchyna_direct_2022} found that galaxies with strong \Civ\ nebular emission such as in System 10 have ultraviolet spectra dominated by stars between 3 and 22 Myr old, with metallicities $\lesssim 10$\% solar.

We verified that the conclusion of a very young and metal-poor stellar population is robust to the assumed stellar IMF and binarity. Repeating the analysis with all IMFs described in \cite{stanway_re-evaluating_2018} gives nearly identical results, including for broken power laws with different high-mass slopes and the \cite{Chabrier:2003ki} IMF, with both binary evolution and single stars only. The best-fit age ranges from 3--5~Myr while the metallicity is consistently $Z=0.001$.

\subsection{Ultraviolet luminosity and star formation rate}
\label{sec:luminosity}

We conduct photometric measurements using the KCWI data, by summing the spectra multiplied by the SDSS $g$ filter throughput curve which is fully covered by the KCWI spectral range. We compare with SDSS photometric measurements of objects in the field of view to determine the zero point. We obtain apparent $g$ magnitudes of 25.4, 25.0, and 25.4 for the northeast, central, and southwest images of System 10 respectively, with an uncertainty of 0.2 magnitudes in the zeropoint. 
From the gravitational lensing model described in \cite{sharma_high_2018}, we obtain magnification factors $\mu = 8.2$, 13.0, and 6.9 for each respective image. 
The intrinsic magnitudes corrected for magnification are in reasonable mutual agreement ($g = 27.7$, 27.8, 27.5) with mean $g = 27.7$. We take the standard deviation of 0.12 magnitudes as an estimate of the combined lens magnification and flux measurement uncertainties, although we caution that there may be additional systematic uncertainty in the magnification. 
Adding the zeropoint uncertainty in quadrature, the intrinsic magnitude of System 10 is $g = 27.7 \pm 0.2$.

The ultraviolet luminosity of System 10, as traced by the observed optical ($g$ band) flux, is far lower than the majority of galaxies studied spectroscopically at similar redshifts. The large samples at $z\sim2$--3 studied by \cite{erb_stellar_2006} and \cite{steidel_structure_2010} have magnitudes $R < 25.5$ (and $g \lesssim 25.5$) with typical $g \simeq 24.5$. System 10 is thus approximately an order of magnitude fainter (i.e., 2--3 magnitudes or a factor of $\sim10\times$ lower luminosity). 
In terms of star formation rate, the observed UV luminosity density corresponds to an unobscured SFR~$= 0.6$~\Msunyr\ based on the \cite{kennicutt_1998} calibration assuming a constant star formation history. This should be considered a lower limit, as a young age and dust attenuation would both cause this SFR to be underestimated. 
Using the lens model, we measure an intrinsic effective radius of $R_e\simeq200$--300 parsecs from a S{\' e}rsic profile fit to source-plane reconstructed NIRC2 images of System 10 (Figure~\ref{f:galaxyimage}). 
This size combined with the SFR limit corresponds to a surface density within the effective radius of $\Sigma_{SFR} \gtrsim 1$~\Msunyr\,kpc$^{-2}$.

\section{Discussion}\label{l:discussion}

The interstellar gas absorption profile of System 10, displaying inflows with no clear outflow signature, is highly unusual in the context of previously studied star forming galaxies at similar redshifts $z\sim2$--3. 
The redshifted (inflowing) ISM absorption is in contrast with the sample of \citet{steidel_structure_2010}, who measured ISM velocity centroids in 89 galaxies at $z\sim2$ using similar methods, finding little evidence for inflowing gas. Likewise \citet{weldon_mosdef-lris_2023} find only 3 robust examples of inflow-dominated ISM absorption in a sample of 134 galaxies at $z\sim2.3$. 
We compare our results with the ISM absorption velocity distribution from the \citet{steidel_structure_2010} and \citet{weldon_mosdef-lris_2023} samples in Figure~\ref{f:steidelsample}, showing that System 10 is a clear outlier with a large positive (inflowing) ISM velocity. 
Moreover, System 10 has an unambiguous inflow signature in the nebular and interstellar component of \Civ, which shows blueshifted emission and redshifted absorption. This is opposite to the common P-Cygni-like profile which indicates outflows (e.g., in the \Lya\ transition; we note that \Lya\ in System 10 is affected by intervening absorption but does show a strong blueshifted emission component visible in Figure~\ref{f:fullspectrum}).

\begin{figure}[htb]
\begin{center}
\scalebox{.75}{\includegraphics{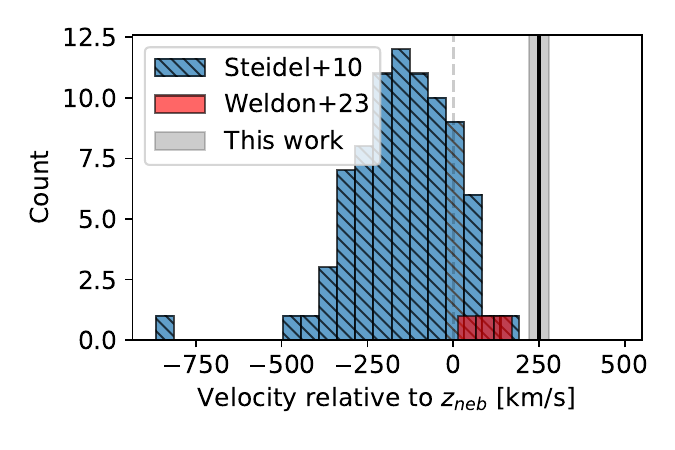}}
\caption{Comparison of the ISM absorption velocity centroid in System 10 with the sample of $z\sim2$ galaxies studied in \cite{steidel_structure_2010} and \cite{erb_stellar_2006}, shown as the blue histogram. Three LRIS-Inflow galaxies studied in \cite{weldon_mosdef-lris_2023} out of a parent sample of 134 are shown as the red histogram.
System 10 is an outlier, with clear inflow indicated by the $+250$~\kms\ velocity (black line, with gray shading showing the $1\sigma$ uncertainty). In contrast, the blue and red histograms show $<200$~\kms\ in all cases, with the vast majority having blueshifted (negative) ISM velocities corresponding to outflows. We note that System 10 has a considerably lower dynamical mass as indicated by its small velocity dispersion and radius, relative to the other samples shown here. We posit that the inflow seen in System 10 is associated with accretion-driven bursty star formation, predicted to occur only in such low-mass galaxies. 
}
\label{f:steidelsample}
\end{center}
\end{figure}

An important aspect of the inflow into System 10 is that it is detected in heavy element transitions (e.g., \Siii, \Civ), indicating a moderate metallicity. Thus we are not witnessing accretion of pristine gas from the intergalactic medium. Instead the accretion likely arises from recycled gas which has been previously ejected via outflows from the galaxy. We cannot rule out accretion from an interacting companion, although we do not detect any companion galaxies within the KCWI field of view, which probes projected distances of $\sim$20--30 kpc in the $z=2.45$ source plane after correcting for the lens magnification. Theoretical work also suggests that accretion of recycling gas dominates over mergers and companions at $z\sim2$, especially at low masses \citep{angles-alcazar_2017}.

The rarity of inflow-dominated ISM absorption signatures suggests that the conditions which create it are short-lived. An intriguing possibility is that we may be witnessing the onset of a bursty star formation episode in a low-mass galaxy. 
Cosmological simulations (e.g., \citealt{governato_2010}; FIRE: \citealt{hopkins_2014}; NIHAO: \citealt{dutton_2016}; APOSTLE and AURIGA: \citealt{bose_2019}) suggest that galaxies at cosmic noon undergo starbursts up to $\lesssim20$\% of the time, with the fraction of stellar mass formed in starbursts being larger for less massive galaxies \citep[e.g.,][]{atek2014}. At low masses $M_{\star}\lesssim10^{9} M_{\odot}$, nearly all stars may be formed during starbursts \citep{sparre_starbursts_2017}. Such simulations show a time sequence whereby gas inflow triggers star formation, and subsequently stellar feedback drives outflows, which remove the gas and cause star formation to cease \citep[e.g.,][]{muratov_gusty_2015,barrow_2020}. The inflow-dominated phase -- before stellar feedback begins driving significant outflows -- is predicted to last only a few Myr. Together with a $\sim100$ Myr period for starbursts \citep[e.g.,][]{muratov_gusty_2015,ting_2024}, this type of inflow-dominated signature may be expected in only $\lesssim$10\% of galaxies at this stellar mass and redshift. 


Our analysis of System 10's ultraviolet spectrum further supports the hypothesis that this is a low-mass galaxy seen shortly after the onset of a starburst. BPASS modeling indicates that the ultraviolet emission is dominated by metal-poor stars (suggesting a low stellar mass; e.g., \citealt{sanders_2015}) with an age of only $\sim$4 Myr. Our measurements in Section~\ref{ss:ismgas} also support a low dynamical mass of a few $\times 10^8~\Msun$.
Cosmological simulations suggest that starburst durations can range from $\sim3$-50 Myr \citep[e.g.,][]{sparre_starbursts_2017}, compatible with our best-fit age. At 4 Myr, supernova feedback has not yet peaked, and mechanical feedback from stellar winds is expected to be weak in such metal-poor systems (thus explaining the relatively weak stellar P-Cygni features in System 10; Figures~\ref{f:fullspectrum}, \ref{f:models}). The young age and low metallicity can thus lead to weak outflows, such that the ISM absorption signatures are dominated by inflowing material which is triggering the starburst. 

Another possibility is that line-of-sight effects may be partly responsible for the dominant inflow signature observed in System 10. 
Outflows may be occurring in a direction away from the line of sight (i.e., not directly between the star forming regions and the observer; \citealt{weldon_mosdef-lris_2023}). Outflowing gas not directly in front of the young stars would not be observable in absorption, as no rest-frame ultraviolet starlight would shine through it. Likewise, geometric effects may also be enhancing the inflow signature observed in System 10, if the inflowing gas is located preferentially along the line of sight. 
The intrinsic size of this galaxy is small compared to the more massive galaxies typically studied at $z\sim2$, with $R_e \simeq 200$-300 pc in near-IR (rest-frame optical) adaptive optics imaging which is sensitive to older stellar populations (Figure~\ref{f:galaxyimage}; Section~\ref{sec:luminosity}).
Such geometric effects can be particularly strong in compact galaxies such as this one. 
Regardless, the lack of outflow signatures in ISM absorption is in stark contrast to samples of more massive galaxies at similar redshift (e.g., Figure~\ref{f:steidelsample}), which display a large outflow covering fraction \citep[e.g.,][]{vasan_2023}. Given that strong outflows are normally expected from the high star formation density in System 10 \citep[$\Sigma_{SFR} \gtrsim 1$~\Msunyr\,kpc$^{-2}$, see Section~\ref{sec:luminosity};][]{heckman_2001,newman_2012}}, we view the covering fraction scenario as a less likely explanation for the lack of outflow signatures.


In summary, the spectroscopic data support a scenario whereby inflow-dominated ISM absorption is arising from bursty star formation in a low mass (and low metallicity) dwarf galaxy, observed only a few Myr after a starburst onset. This behavior is common in FIRE cosmological simulations of low mass galaxies at the redshift $z\sim2.5$ of System 10 \citep{sparre_starbursts_2017,muratov_gusty_2015}, with a timescale compatible with the young stellar population age found here (Section~\ref{ss:stellarmodels}). The lack of outflowing gas is explained by the young age, although geometric effects cannot be ruled out. 
Based on the starburst timescale and duty cycle in cosmological simulations, we may expect such an inflow-dominated signature up to $\sim$10\% of the time in low-mass galaxies. However, such galaxies are most easily observed during starburst phases when their rest-frame ultraviolet luminosity peaks. If this explanation is correct, then we expect inflow signatures to be relatively common among $z\gtrsim2$ dwarf starburst galaxies, particularly those with young starburst ages as inferred here from the ultraviolet stellar spectrum. Gravitational lensing magnification offers a pragmatic approach to characterize these intrinsically faint galaxies, and indeed the results herein are only possible thanks to the combination of strong lensing and deep exposures.

\section{Conclusions}\label{l:conclusions}

In this work we present deep spectroscopic observations of an intrinsically low-luminosity galaxy at redshift $z=2.45$, from an 8-hour integration with Keck/KCWI aided by strong gravitational lensing magnification. The intrinsic luminosity, nebular kinematics, and stellar spectra indicate that the target is a low-mass and low-metallicity (i.e., dwarf) galaxy with rest-UV emission dominated by a very young ($\sim$4 Myr) stellar population. 
Analysis of the nebular \Civ\ emission and several prominent interstellar absorption lines reveal an unambiguous signature of inflowing gas, with a radial velocity of $250\pm30$ (stat.) $\pm35$ (sys.)~\kms. In contrast to previous spectroscopic samples of more massive star-forming galaxies at similar redshift, we do not see strong signatures of outflows in the interstellar absorption. We propose that the strong inflow and lack of outflow signatures in this galaxy are a consequence of bursty star formation, possibly combined with geometric effects from the small size and line-of-sight viewing angle. 
We note that the inflowing gas is observed in heavy element transitions, such that it is not pristine intergalactic gas, but rather has been enriched by previous star formation. 

Cosmological simulations support the picture of inflow-dominated ISM kinematics during the onset of a starburst in low-mass galaxies at $z\gtrsim2$ \citep[e.g.,][]{muratov_gusty_2015,sparre_starbursts_2017}. This phase is expected to be short-lived, which partly explains the rarity of this observational signature. Much of the inflowing material is predicted to be recycled from previous outflows \citep[e.g.,][]{angles-alcazar_2017,ford_2014}, as required for the inflow signature we observe here in heavy element transitions. 
To our knowledge, this work is the first observational verification of the inflow-dominated phase predicted to be common for dwarf galaxies at such redshifts. 
The burstiness of star formation required for this type of signature is uncommon among higher mass galaxies (stellar mass $\gtrsim 10^9 \Msun$), which make up the vast majority of spectroscopic observations at cosmic noon. If a survey of similarly low-mass galaxies at cosmic noon were to be conducted, we predict that of order 10\% might show inflow-dominated ISM absorption profiles -- with a higher fraction among those with young starburst ages. Despite their low luminosity, this work demonstrates the feasibility of deep rest-UV spectroscopy (e.g., 
combined with strong gravitational lensing or with future extremely large telescopes) to study the nature of bursty star formation and the baryon cycle in low-mass galaxies. 


\begin{acknowledgments}
We thank Soniya Sharma and Johan Richard for helpful discussions and for providing the gravitational lensing model. 
TJ, KVGC, and SR gratefully acknowledge financial support from the National Science Foundation (NSF) through grant AST-2108515, the Gordon and Betty Moore Foundation through Grant GBMF8549, the National Aeronautics and Space Administration (NASA) through grant HST-GO-16773, and from a Dean's Faculty Fellowship. 
RSE acknowledges generous financial support from the Peter and Patricia Gruber Foundation. 
KG and GK acknowledge the support of the Australian Research Council Centre of Excellence for All Sky Astrophysics in 3 Dimensions (ASTRO 3D), through project number CE170100013. 
Funding for this research was provided by the NSF's physics REU program at the University of California, Davis under grant PHY-2150515. This research has made use of the Keck Observatory Archive (KOA), which is operated by the W.~M. Keck Observatory and the NASA Exoplanet Science Institute (NExScI), under contract with the National Aeronautics and Space Administration. The data presented herein were obtained at the W.~M. Keck Observatory, which is operated as a scientific partnership among the California Institute of Technology, the University of California and the National Aeronautics and Space Administration. The Observatory was made possible by the generous financial support of the W. M. Keck Foundation. The authors wish to recognize and acknowledge the very significant cultural role and reverence that the summit of Maunakea has always had within the indigenous Hawaiian community. We are most fortunate to have the opportunity to conduct observations from this mountain. 
\end{acknowledgments}

%

\vspace{5mm}


\bibliography{bibliography}
\bibliographystyle{aasjournal}

\end{CJK*}
\end{document}